\documentclass[conference]{IEEEtran}
\usepackage[utf8]{inputenc}

\usepackage{tabu}
\usepackage{enumitem}
\usepackage[fleqn]{mathtools}
\usepackage{url}
\usepackage{float}
\usepackage{tabularx}
\usepackage[bottom]{footmisc}
\usepackage[hpos=0.7\paperwidth, % .98 to prevent bleed
            vpos=0.04\paperwidth,
            angle=90]{draftwatermark}

\title{An Experimental Investigation of Tuning QUIC-Based Publish-Subscribe Architectures in IoT\vspace{-3mm}}

\author{
  \IEEEauthorblockN{
    Darius Saif\IEEEauthorrefmark{1}, Ashraf Matrawy\IEEEauthorrefmark{2}
  }
  \IEEEauthorblockA{
    Carleton University, Department of Systems and Computer Engineering\IEEEauthorrefmark{1}, School of Information Technology\IEEEauthorrefmark{2}\\
    Email: {\{dariussaif\IEEEauthorrefmark{1}, amatrawy\IEEEauthorrefmark{2}}\}@sce.carleton.ca 
  }
}

\begin{document}
\maketitle

\SetWatermarkText{Authors' draft for soliciting feedback: \today}
\SetWatermarkColor[gray]{0.5}
\SetWatermarkFontSize{0.6cm}
\SetWatermarkAngle{0}
\SetWatermarkHorCenter{11cm}

%%%%%%%%%%%%%%%%%%%%%%%%%%%%%%%%%%%%%%%%%%%%%%%%%%%%%%
\begin{abstract}

There has been growing interest in using the QUIC transport protocol for the Internet of Things (IoT). In lossy and high latency networks, QUIC outperforms TCP and TLS. Since IoT greatly differs from traditional networks in terms of architecture and resources, IoT specific parameter tuning has proven to be of significance. While RFC 9006 offers a guideline for tuning TCP within IoT, we have not found an equivalent for QUIC. This paper is the first of our knowledge to contribute empirically based insights towards tuning QUIC for IoT. We improved our pure HTTP/3 publish-subscribe architecture and rigorously benchmarked it against an alternative: MQTT-over-QUIC. To investigate the impact of transport-layer parameters, we ran both applications on Raspberry Pi Zero hardware. Eight metrics were collected while emulating different network conditions and message payloads. We enumerate the points we experimentally identified (notably, relating to authentication, MAX\_STREAM messages, and timers) and elaborate on how they can be tuned to improve resource consumption and performance. Our application offered lower latency than MQTT-over-QUIC with slightly higher resource consumption, making it preferable for reliable time-sensitive dissemination of information.
\end{abstract}

\begin{IEEEkeywords}
IoT, QUIC, HTTP/3, MQTT
\end{IEEEkeywords}

\IEEEpeerreviewmaketitle

%%%%%%%%%%%%%%%%%%%%%%%%%%%%%%%%%%%%%%%%%%%%%%%%%%%%%%
\section{Introduction}
The Internet of Things (IoT) is a paradigm shift where objects can autonomously sense and control our environment. Unlike in traditional networks, many embedded systems exchange data Machine-to-Machine (M2M) with little direct intervention from a human operator. Form factor and unit cost restrictions typically impose practical limitations on a node's processing power, battery, and memory \cite{rfc7228}. The networks over which these devices operate can also be lossy and have unpredictable levels of bandwidth -- underscoring a fundamental need for reliable yet lightweight communication.

Because of IoT's key distinctions, many traditional technologies and protocols require special consideration when used; especially in the transport layer. For example, RFC 9006 \cite{rfc9006} offers a guideline for tuning TCP in IoT. It cites TCP's weaknesses in IoT as long header lengths, unsuitability for multi-cast, and always-confirmed data delivery.

QUIC (not an acronym) is a recent general-purpose transport protocol which was designed to address certain deficiencies of TCP \cite{rfc9000}. QUIC fares better than TCP and TLS in networks with high loss, Round-Trip Time (RTT), and low bandwidth \cite{iccPaper} -- which are typical conditions in IoT. The major advantages QUIC offers include: i.) integrated security, ii.) reduced connection establishment overhead (0-RTT), iii.) variable-length integer encoding, iv.) connection migration for mobility, and v.) a user-space implementation not dependent on a system's kernel. Additionally, QUIC brings the notion of streams and multiplexing to the transport layer in order to reduce the effects of Head-of-Line-Blocking (HoLB).

For these reasons, QUIC has been proposed as a candidate transport layer solution within IoT \cite{eggert2020towards}. While many of QUIC's features may be desirable, Langley \textit{et al.} \cite{langley2017quic} note the original implementation of QUIC was built for \textit{"rapid feature development and ease of debugging, not CPU efficiency"}.

The aim of this paper is to provide empirically based insights towards tuning QUIC in IoT for better performance and resource management. To this end, we improved upon our HTTP/3 publish-subscribe (pub-sub) architecture (henceforth called H3) \cite{cscnpaper} and rigorously benchmarked it against MQTT-over-QUIC \cite{fernandez2020and}. We ran both applications on Raspberry Pi Zero hardware while emulating different network conditions and message payloads. We collected 8 metrics from our real traffic test cases and performed root cause analysis on the data. From this experimentation, we i.) compiled set of guidelines for tuning QUIC for IoT, similar to TCP's RFC 9006 and ii.) found H3 to have lower latency but a slightly larger footprint. 

\textbf{Contributions} of this paper include: i.) first technical paper of our knowledge to experimentally tune QUIC for use within IoT (Section V), ii.) improved implementation of our pure H3 pub-sub application (Section III), iii.) proposal of a high-watermark scheme for sending MAX\_STREAM advertisement frames, still adhering to QUIC RFC 9000 (Section V), and iv.) proposal of connection-level basic authentication (basicAuth) for H3 triggered by any QUIC stream (Section IV).

The rest of this paper is organized as follows: Section II discusses related work integrating QUIC into IoT. Our pub-sub implementation and MQTT-over-QUIC are described in Sections III and IV. QUIC parameters we experimentally tuned are discussed in Section V. Sections VI and VII cover our experimental setup and evaluation methodology. The results of our experiments are provided in Section VIII. Finally, conclusions are drawn in Section IX. 

%%%%%%%%%%%%%%%%%%%%%%%%%%%%%%%%%%%%%%%%%%%%%%%%%%%%%%
\section{Related Work}
Eggert showed that certain classes of IoT devices are capable of running QUIC \cite{eggert2020towards}. Two 32-bit microprocessor development boards (Particle Argon \& ESP32-DevKitC V4) were installed with \textit{Quant} and \textit{picoTLS} in order to support QUIC. Storage space, battery consumption, as well as memory and CPU usage on the boards were monitored while 5KB objects were downloaded. It was found that the QUIC implementation required about 63KB of flash, 16KB of heap memory, 4KB of stack memory, and 0.9J energy per transaction. This was deemed sufficient and that, with further optimizations, QUIC could run on 16-bit processors as well. Unlike our work, HTTP/3 was not included in Eggert's network stack.

Kumar and Dezfouli \cite{kumar2019implementation} outfitted Chromium's Google QUIC (GQUIC) to carry MQTT. Because MQTT and GQUIC both run in user-space, they had to write inter-process communication APIs and redesign various data structures. Raspberry Pi 3 Model Bs were used for each network entity in their setup. Three network categories were used in their testing: wired, wireless, and long distance. During connection establishment, MQTT-over-GQUIC reduced packets exchanged with the broker by up to 56.25\%. To test HoLB, packets were randomly dropped. MQTT (over TCP) was shown to have higher latency in every network configuration. Memory and processor utilization of the broker were also monitored for half-open connections. Although MQTT-over-GQUIC used slightly more resources, it relinquished up to 83.24\% and 50.32\% more processor and memory usage respectively compared to MQTT. Lastly, MQTT-over-GQUIC was found to be more resilient in cases where the broker's IP address changed mid-transaction.

Fernandez \textit{et al.} \cite{fernandez2020and} had implemented MQTT over QUIC (draft version 27) and compared it against TCP-based MQTT. Their implementation was made available open-source. They used NS-3 and Linux containers to emulate WiFi, 4G, and satellite network profiles. On each profile, single and multiple MQTT connections were tested. For single connections, 1000 messages were published and the time delta from starting transmission to the subscriber receiving the messages was recorded. They found that QUIC outperformed TCP in every configuration, by up to 40\%. In the multiple connection testing, they focused on connection establishment time. After publishing one message, the connection to the publisher and broker was closed. Afterwards, another message was published on a new connection. QUIC proved to perform slightly better than TCP and it exhibited less variation. Recently, the authors positioned their work for Industrial IoT \cite{s21175737} and have also extended their setup to include IoT hardware devices \cite{s22103694}.

Efforts to integrate QUIC with Constrained Application Protocol (CoAP) \cite{herrero2021analysis} and Advanced Message Queuing Protocol (AMQP) \cite{IQBAL2023109640} have also taken place with favorable results for QUIC-based transport compared to TCP and UDP. 

Rather than integrating an existing protocol with QUIC, we have extended our HTTP/3 pub-sub implementation \cite{cscnpaper}. We hypothesized HTTP/3 would make better use of QUIC, as it is HTTP/3's native transport protocol. Our implementation does not require any conversion code or changes to data structures, unlike the above works. We discuss low-level tuning of our underlying QUIC library’s parameters for effective use in IoT while past works do not. We use hardware devices to test our implementation against MQTT-over-QUIC \cite{fernandez2020and} and include additional test-cases and metrics not considered in Fernandez \textit{et al.}'s study. Namely, we collect data from MQTT-over-QUIC in 2 modes of Quality of Service (QoS): Fire-and-Forget and Acknowledged Delivery. The full list of improvements to our pub-sub application and test apparatus is shown in Table \ref{tab:h3enhancements}.

{
\tabulinesep=1mm
\begin{table}[!htb]
  \centering
  \caption{Summary of Extensions over our Past Work \cite{cscnpaper}}
  \begin{tabu} to 0.48\textwidth {|X[3.3]|X[1.7]|X[2.7]|}
     \hline
      
      &\centering \textbf{ Past Work }&
      \centering \textbf{This Work}\\\hline
      
      \textbf{Testbed Environment}&
      Ubuntu VMs&
      Raspberry Pi Zeros\newline Ubuntu VMs\\\hline
      
      \textbf{Device Overhead Metrics}&
      CPU\newline Memory&
      CPU\newline Memory\newline Power Consumption\\\hline

      \textbf{QUIC Version}&
      draft-29&
      RFC 9000\\\hline
      
      \textbf{QUIC Tuning}&
      Defaults&
      Tuned for IoT\\\hline

      \textbf{Statistical Representation}&
      Mean&
      Box \& Whisker Plots\\\hline
      
      \textbf{Tuning Variables}&
      Message Size&
      Message Size\newline Packet Loss \%\newline Number of Messages\newline RTT\\\hline
      
      \textbf{MQTT Modes \newline Tested}&
      Fire \& forget&
      Fire \& forget \newline ACK'ed Delivery\\\hline

 \end{tabu}
  \label{tab:h3enhancements}
\end{table}
}

%%%%%%%%%%%%%%%%%%%%%%%%%%%%%%%%%%%%%%%%%%%%%%%%%%%%%%
\section{Publish-Subscribe Architectures for IoT}

\begin{figure}[b]
\centering
\vspace{-2mm}
\includegraphics[width=1.9in]{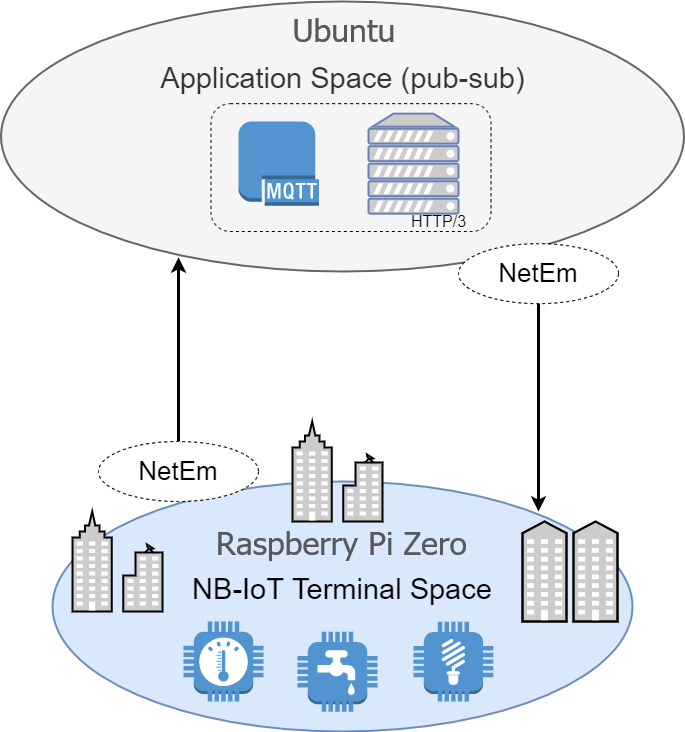}
\vspace{-0.5mm}
\caption{Emulated Network Topology}
\label{nbiot}
\end{figure}

As a widely used application in IoT, we focus on M2M communication in our investigation of tuning QUIC \cite{kumar2019implementation}. Our pub-sub implementation, H3, is compared against the state-of-the-art MQTT-over-QUIC as a baseline. A high-level view of these architectures is illustrated in Figure \ref{nbiot}.

\subsection{MQTT-over-QUIC}
MQTT organizes information into topics and introduces a broker entity. The broker manages logistics of authentication as well as accepting and forwarding information -- alleviating the strain on publishing and subscribing clients. Fernandez \textit{et al.}'s MQTT-over-QUIC (henceforth called MQ) used QUIC-GO \cite{lucasClem}, a pure GO implementation of QUIC. They had integrated the QUIC-GO library with Eclipse Paho\footnote{https://github.com/pgOrtiz90/paho.mqtt.golang} and VolantMQ\footnote{https://github.com/fatimafp95/volantmq\_2}. These packages were used to realize publishers/subscribers and the broker, respectively. In their implementation, both versions 3.1 and 3.1.1 of MQTT were supported. We have upgraded their open-source code to make use of a newer release version of QUIC-GO which supports RFC 9000.

\subsection{HTTP/3-based Publish-Subscribe}
\textbf{Overview:} Our H3 architecture was adapted from Drift: an open-source GO implementation \cite{technosophos}. We retooled it to reap the many overhead reductions offered by HTTP/3, compared to other HTTP versions. QUIC-GO was also leveraged in our implementation, for parity with MQ. Unlike MQTT, our architecture does not support a fire-and-forget QoS type. Rather, acknowledged delivery is solely supported.

We scrapped the entirety of Drift’s \textit{transport} package (which implemented HTTP/2 \cite{bradfitz}) and wrote an equivalent one with QUIC-GO’s HTTP/3 stack. These changes posed consequences in Drift’s \textit{client} package, housing the methods used for mapping pub-sub concepts to HTTP semantics, which are shown in Table \ref{tab:h3pubsubsemantics}. In each case, the client encodes the \textit{topic} name into the request header. We also upgraded the \textit{server} package by calling QUIC-GO's HTTP/3 listener.

{
\tabulinesep=1mm
\begin{table}[!t]
  \centering
  \caption{H3 Publish-Subscribe Semantic Mappings}
  \begin{tabu} to 0.48\textwidth {|X[1.5]|X[1.1]|X[4.7]|}
     \hline
      
      Topic Exists&HEAD&Server returns 200 if it exists, else 404.\\\hline
      
      Create Topic&PUT&Server sanity checks the topic name and creates the topic if checks pass.\\\hline
      Delete Topic&DELETE&Server cleans up topic data and unsubscribes all parties from the topic.\\\hline
      
      Publish \newline to Topic&POST&Client encodes topic name in URL and includes data in message body, server stores information and pushes to subscribers.\\\hline
      
      Subscribe \newline to Topic&GET&Client spawns a listener for incoming events, server appends node to the list of nodes subscribed to the topic.\\\hline

 \end{tabu}
  \label{tab:h3pubsubsemantics}
  \vspace{-2mm}
\end{table}
}

\section{Authentication Analysis for H3}
In this section, we discuss the impact of using basicAuth \cite{rfc7235} instead of X.509 certificates in our H3 pub-sub code.

A GO profiler, \textit{pprof}, was used to investigate memory and CPU of both MQ and H3 in our previous work \cite{cscnpaper}. \textit{pprof}'s \textit{MemProfileRate} was driven to 0 in order to record all allocated memory blocks. The profiler collected client-side data from connection establishment, publishing messages, and connection termination. Then, the \textit{top} command was used to identify the 5 most memory-hungry functions of each application. We found that H3 used approximately 10 times more memory than MQ. The cause of this disparity was because our implementation used X.509 certificates while MQ did not. After modifying our H3 pub-sub code to use basicAuth, both H3 and MQ employed credential-based authentication.

Table \ref{tab:topmem509} lists H3's top 5 memory consumers when X.509 certificates were used. The \textit{flat} columns represent memory that was allocated (and held) by that function, whereas \textit{cum} columns are inclusive of the function's children. The \textit{sum\%} column is the summation of the current, and all previous, \textit{flat\%} values. The methods listed all come from standard GO libraries, whose documentation and source code are available online\footnote{https://pkg.go.dev/std}. All of the methods are directly involved in parsing, encoding, and decoding operations for the X.509 certificate.

\begin{table*}
  \centering
  \caption{Top 5 Memory Consumers for H3 with X.509}
  \vspace{-2mm}  
  \begin{tabular}{p{0.07\linewidth}p{0.05\linewidth}p{0.05\linewidth}p{0.07\linewidth}p{0.06\linewidth}p{0.18\linewidth}p{0.34\linewidth}}   
    flat&flat\%&sum\%&cum&cum\%&Method&Description\\\hline  
    361.78kB&12.51\%&12.51\%&1193.60kB&41.26\%&crypto/x509.parseCertificate&Parses a single certificate from encoded ASN.1 data\\\hline

    313.75kB&10.84\%&23.35\%&313.75kB&10.84\%&bytes.makeSlice&Creation of byte slices for parsing routines\\\hline
      
    265.48kB&9.18\%&32.53\%&265.48kB&9.18\%&reflect.unsafe\_NewArray&Creation of an array, used by bytes.makeSlice\\\hline
    244.89kB&8.46\%&40.99\%&1319.42kB&45.61\%&encoding/asn1.parseField&Parse an ASN.1 value, given a byte slice and offset\\\hline

    229.84kB&7.94\%&48.94\%&263.97kB&9.12\%&encoding/pem.Decode&Returns the next PEM block and the input's remainder\\\hline
  \end{tabular}
  \vspace{-2mm}  
  \label{tab:topmem509}
\end{table*}

With basicAuth, credentials must be included in the header of the client's HTTP request. If the credentials are known to the server, the client is authenticated and the request will be processed. If not, an HTTP 401 status code of Unauthorized is returned to the client and the request is not processed, meaning the client may not publish or subscribe to any topics.

Table \ref{tab:topmemBA} lists H3's top 5 memory consumers when basicAuth is used. In total, a memory savings of 2.7MB was achieved compared to when X.509 was in use. CPU profiling showed a utilization reduction of 37.5\% was yielded as well. For an IoT environment, these savings are of great importance, as the devices are typically resource constrained.

\begin{table*}
  \centering
  \caption{Top 5 Memory Consumers for H3 with basicAuth}
  \vspace{-2mm}
  \begin{tabular}{p{0.07\linewidth}p{0.05\linewidth}p{0.05\linewidth}p{0.07\linewidth}p{0.06\linewidth}p{0.545\linewidth}}  
    flat&flat\%&sum\%&cum&cum\%&Method\\\hline  
    24.25kB&11.44\%&11.44\%&24.25kB&11.44\%&github.com/lucas-clemente/quic-go.init.0.func1\\\hline

    18.38kB&8.66\%&20.10\%&18.38kB&8.66\%&github.com/lucas-clemete/quic-go/internal/wire.init.0.func1\\\hline
      
    14.73kB&6.95\%&27.05\%&22.05kB&10.40\%&crypto/sha256.(*digest).Sum\\\hline
    11.59kB&5.47\%&32.52\%&11.59kB&5.47\%&crypto/sha256.New\\\hline

    10.50kB&4.95\%&37.47\%&10.50kB&4.95\%&encoding/bytes.makeSlice\\\hline
  \end{tabular}
  \label{tab:topmemBA}
  \vspace{-2.5mm}
\end{table*}

We note the change to basicAuth as a security-performance trade-off. Client-end X.509 authentication is not required by TLS's standard \cite{rfc8446} and not using it yielded sizable resource consumption reductions. On one hand, basicAuth is lightweight but it requires credential distribution and management. While brute forcing credentials \textit{is} a possibility with basicAuth, policies can be applied to blacklist clients with too many failed login attempts. On the other hand, X.509 certificates offer trusted public-keys at the price of high resource consumption. Managing such certificates has proven to be a challenge in practice however \cite{sslLandscape}, especially for IoT.

\textbf{Further Reducing basicAuth Overhead:} As an additional contribution, we designed a basicAuth scheme operating on the QUIC connection level for our H3 client and server. That is, if a QUIC stream \textit{S1} is authenticated by the server on a connection \textit{C}, subsequent streams will also be considered authenticated for the life of \textit{C}. The client reads the server's HTTP response codes to determine whether its connection is authenticated. This way, credentials need not be provided in every single request header -- amounting to less device processing and fewer bytes transmitted.

We implemented basicAuth for the client by encoding its credentials into the HTTP request header using \texttt{SetBasicAuth()}. On the server, we authenticated client credentials by calling \texttt{BasicAuth()}. Both functions are a part of GO's \texttt{net/http} package. Lastly, we added a flag to QUIC-GO's \texttt{session} structure for the server to determine the authentication status of the client connection. If the flag was driven to true for a unique QUIC connection ID, no further authentication was performed. This performance enhancement is made possible by disabling QUIC's 0-RTT in our setup. RFC 9001 \cite{rfc9001} states: \textit{"Disabling 0-RTT entirely is the most effective defense against replay attack"}. Beyond this, we have not performed a full security analysis on our enhancement. 

%%%%%%%%%%%%%%%%%%%%%%%%%%%%%%%%%%%%%%%%%%%%%%%%%%%%%%
\section{QUIC Implementation \& IoT Tuning}

In this section, we discuss the details of the QUIC library we used along with the various features enabled in our setup. Our contributions towards tuning QUIC for effective and resource-friendly operation in IoT are also described and justified. Identifying and tuning these portions of code was done as an iterative process with the help of code profiling, network captures, and qlogs \cite{qlog}. The changes discussed affect both MQ and H3 unless specified otherwise.

In our setup, QUIC-GO \cite{lucasClem} release v0.25.0 powered \textit{both} the H3 and MQ implementations. The library adhered to the QUIC standard RFC 9000 \cite{rfc9000}. QUIC-GO did not provide support for Unreliable Datagram Extensions, as defined in RFC 9221 \cite{rfc9221}, and it was therefore not used in our setup.

All the \textit{quic\_transport\_parameters} of H3 and MQ were tuned to be identical to one another. Both used the \texttt{TLS\_CHACHA20\_POLY1305\_SHA256} TLS cipher suite, CUBIC congestion control \cite{ha2008cubic}, path MTU discovery, and basicAuth. Dynamic table QPACK was not used for H3; only the static table and Huffman encoded string literals were employed. QPACK \cite{rfc9204} is HTTP/3's method of header compression, which is resilient against out-of-order delivery.

\subsubsection{\textbf{MAX\_STREAM Handling}}
When QUIC stream IDs are consumed and retired, it is up to the server to inform the client as to how many new streams can be opened. The behavior of the QUIC-GO library is to update the MAX\_STREAM count on every closed stream. According to RFC 9000 \cite{rfc9000}: \textit{“As with stream and connection flow control, this document leaves implementations to decide when and how many streams should be advertised to a peer via MAX\_STREAMS. Implementations might choose to increase limits as streams are closed, to keep the number of streams available to peers roughly consistent.”} 

A point of our contribution includes design of a high-watermark scheme whereby the server will inform the client once 50\% of the originally negotiated streams have been closed. This reduced the number of MAX\_STREAM frames sent, saving network resources and reducing signaling on the constrained IoT device. It is assumed that the stream consumption rate for most IoT publisher devices will not warrant a 1:1 MAX\_STREAM frame to stream close ratio.

\subsubsection{\textbf{QUIC Timers}}
Due to the higher RTT typically associated with various communication standards for IoT, several QUIC-GO timers were increased to be reflective of this. The \textit{DefaultHandshakeTimeout}, \textit{DefaultHandshakeIdleTimeout}, and \textit{MinRemoteIdleTimeout} were all tuned to be more accommodating to peers with higher RTTs. This ensured that a peer would not close a connection after a period of inactivity.

\subsubsection{\textbf{ACK Range Handling}}
RFC 9006 \cite{rfc9006} states that: \textit{“If a device with less severe memory and processing constraints can afford advertising a TCP window size of several MSSs, it makes sense to support the SACK option to improve performance… a sender (having previously sent the SACK-Permitted option) can avoid performing unnecessary retransmissions, saving energy and bandwidth, as well as reducing latency”}. QUIC's mode of acknowledgements is similar in principle to TCP Selective Acknowledgements (SACKs), as it uses \textit{Range} and \textit{Gap} fields to efficiently specify what has been received and what has not \cite{rfc9000}. As per RFC 9000, a \textit{max\_ack\_delay} is negotiated at the start of a connection to declare the maximum time an endpoint can take before responding to an ACK eliciting frame. We increased this value from 25ms to $0.5*RTT$ in order to make more efficient use of ACK ranges.

\subsubsection{\textbf{Default Initial RTT}}
We previously observed that the client would send 7 \textit{Initial} frames by the time a response from the broker node was received \cite{cscnpaper}. This meant needless energy consumption on the device as well as wasted network resources. Again, this behavior was attributed to the high RTT associated with the IoT network profile emulated. When a QUIC connection is being established, neither endpoint has knowledge of the delay to the other endpoint; \textit{defaultInitialRTT} is used in the QUIC-GO code to provide an initial guess. QUIC-GO's default behavior is to wait up to 2*\textit{defaultInitialRTT} to retransmit an \textit{Initial} frame. In the context of resource-constrained IoT, the accuracy of this initial guess is important: if too low, resources are gratuitously consumed, as observed in this case. If too high, the client will allow too wide a grace period if the \textit{Initial} frame is indeed lost. We have changed \textit{defaultInitialRTT} from 100ms to 1500ms.

\subsubsection{\textbf{H3 RequestWriter Buffer}}
With \textit{pprof} profiling, we found that the HTTP/3 module of the QUIC-GO library would allocate a fixed size buffer of 8KB for writing client requests. Before factoring in GO's memory garbage collection, this can amount to high client-end resource consumption per request. In the context of a resource-constrained environment, optimizations can be made in this regard. We operate under the assumption that devices using pub-sub in IoT will typically produce rather small requests (like publishing temperature data from a sensor) which will not come close to the pre-allocated 8KB. Given this assumption, we modified the code to take a more dynamic approach in allocating the buffer.

%%%%%%%%%%%%%%%%%%%%%%%%%%%%%%%%%%%%%%%%%%%%%%%%%%%%%%
\section{Experimental System Setup}

A hybrid environment of hardware devices and Ubuntu 18.04.4 (kernel 5.4.0-109) Virtual Machines (VMs) was used in our test setup. This design choice was made in order to scale the network size (i.e. introduce additional publishing or subscribing nodes) if needed. A Lenovo T490 PC running 64-bit Windows 10 with 24GB RAM and an Intel i5-8365U CPU hosted the VMs. VMs were allocated 1 processor core and 1024 MB of memory. There were two Raspberry Pi Zero boards whose specifications are listed in Table \ref{tab:rPi}. The Raspberry Pi Zero represented a low-cost and suitably constrained device, albeit powerful enough to run ARM Linux.

{
\tabulinesep=1mm
\begin{table}[!b]
  \vspace{-2.5mm}
  \caption{Raspberry Pi Zero Specifications}
  \centering
  \begin{tabu} to 0.475\textwidth {|X[1.8]|X[4]|}
     \hline
      
      CPU&Single core, 1 GHz\\\hline

      Architecture & ARM\\\hline
      
      Memory&512 MB\\\hline
      
      Operating System & Raspbian 11 (bullseye) 5.10.92+ \#1514\\\hline
      
 \end{tabu}
  \label{tab:rPi}
\end{table}
}

Raspberry Pi hardware nodes were configured as publishers. All data presented in Section VIII was collected solely from the hardware nodes in our setup. The broker was realized as a VM -- typically, a broker will be a higher-powered node compared to publisher and subscriber entities, meaning it did not need to be modelled as a constrained device. 

We deemed it imperative to perform our tests in a controlled network environment. The reason for this was to minimize any external factors which could cause fluctuations in the data that we collected. This helped to ensure that i.) only the effects of our parameter tuning had a bearing on the results and ii.) outlier data points could be better understood.

\textbf{Network Environment \& Settings:} All devices were networked through a Technicolor CGM4331COM router over 2.4GHz 802.11n WiFi. Without any network modelling (which is discussed in Section VII), the average ping time and \textit{iperf} reported average link speed from publisher to broker were 12.24 ms and 24.3 Mbits/sec, respectively. Each entity used a Maximum Transmission Unit (MTU) of 1500 bytes.

%%%%%%%%%%%%%%%%%%%%%%%%%%%%%%%%%%%%%%%%%%%%%%%%%%%%%%
\section{Evaluation Approach}
%%%%%%%%%%%%%%%%%%%%%%%%%%%%%%%%%%%%%%%%%%%%%%%%%%%%%%
\subsection{Network Modelling}

NetEm \cite{hemminger2005network} was leveraged to emulate realistic resource-constrained IoT network conditions. Delay and rate throttling impairments were applied to outgoing packets on the publisher and broker network interfaces, as shown in Figure \ref{nbiot}. As other works \cite{fernandez2020and, s21175737, 9854951} have shown, emulation is effective in producing network conditions reflective of IoT. We combined such emulation with the use of constrained hardware devices to accurately model our intended case study.

Network parameters of 3GPP's second generation of Narrowband-IoT (NB-IoT) \cite{zayas20173gpp} were used in this paper. Given NB-IoT's lower cost and minimal battery usage, it is a suitable choice for resource-constrained IoT applications not requiring mobility \cite{zayas20173gpp}. These parameters, featuring modest data rates and high latency, are summarized in Table \ref{tab:nbiot}.

{
\tabulinesep=1mm
\begin{table}[!t]
  \caption{NB-IoT CAT NB2 Network Parameters}
  \centering
  \begin{tabu} to 0.475\textwidth {|X[1.8]|X[4]|}
     \hline
      
      Downlink Rate&127 kbit/s (broker interface)\\\hline
      
      Uplink Rate&159 kbit/s (publisher interface)\\\hline
      
      Round-Trip Time&2 seconds (split between broker and publisher)\\\hline
      
 \end{tabu}
  \smallskip
  \label{tab:nbiot}
  \vspace{-3mm}
\end{table}
}

\subsection{Metrics Collected \& Statistical Representation}
All experiments were run for 50 iterations in order to gauge consistency of the data. For analytical purposes, a packet capture was taken during each iteration using tshark. QUIC specific event reporting from qlog \cite{qlog} was enabled on the client as well. We investigated our H3 implementation against two modes of QoS for MQ: Fire-and-Forget (MQ-FF) and Acknowledged Delivery (MQ-AD). We sought to quantify the resource footprint of each application on the constrained devices while also giving weight to time-sensitivity and performance. Thus, we collected the metrics shown in Table \ref{tab:metrics}. 

{
\tabulinesep=1mm
\begin{table}[!b]
  \vspace{-4mm}
  \caption{Summary of Metrics Collected}
  \centering
  \begin{tabu} to 0.475\textwidth {|X[1.5]|X[4]|}
     \hline
      
      Execution Time&
      The time from the transaction's start to finish\\\hline
      
      Time to First Data Frame (TtFDF)&
      The time between the publisher's \textit{Initial} frame and its first data frame (H3 $DATA\_FRAME$ or MQ $PUBLISH$ packet, respectively)\\\hline
      
      CPU Consumption&
      The aggregate time the program occupied the device's CPU, as reported by \textit{pprof}\\\hline
      
      Memory Consumption&
      The aggregate bytes of memory required for the program's execution, as reported by \textit{pprof} (not factoring in GO's garbage collection)\\\hline
      
      Power \newline Consumption&
      The average power for all process IDs belonging to the program, as reported by Linux's PowerTop\\\hline
      
      Bytes \newline Transmitted&
      The total number of bytes transmitted between the two endpoints, as reported by tshark captures\\\hline
      
      Packets \newline Transmitted&
      The total number of packets transmitted between the two endpoints, as reported by tshark captures\\\hline
      
      Program \newline Footprint&
      The total number of bytes the compiled QUIC-powered architecture occupied on the device\\\hline 
      
 \end{tabu}
  \smallskip
  \label{tab:metrics}
\end{table}
}

Whisker plots were generated using each iteration's data for statistical reporting purposes. The whiskers represent the minimum and maximum values of the Inter-Quartile Range (IQR). The top and bottom ends of the box indicate the median values of the 3rd and 1st quartiles, respectively. The line splitting the box indicates the median. Mean values are shown with an 'X' symbol and outliers with an 'O'.

%%%%%%%%%%%%%%%%%%%%%%%%%%%%%%%%%%%%%%%%%%%%%%%%%%%%%%
\section{Results}

The H3 and MQ server code was compiled on GO v1.16.15 with a Linux OS type and \textit{amd64} architecture. Their respective program footprints amounted to 12MB and 15MB. Client programs were compiled for Linux and an \textit{ARM} architecture type for the Raspberry Pi Zero. The average program footprint was 8.2MB and 8.1MB for H3 and MQ, respectively.

As indicated in Table \ref{tab:h3enhancements}, a total of 4 experiment classes were conducted to comprehensively investigate different behaviors of the tuned QUIC applications: message scaling, RTT effects, message payload size, and packet loss effects.

\subsection{Load Scaling (Number of Publish Messages)}

In these tests, 32-byte messages were sent to a topic on the broker with the NB-IoT profile in place. The purpose of this was to give a glimpse into each architecture under sustained load. The number of messages sent over the QUIC connection range from 1, 10, 25, 50, and 100 for each. The publish message inter-arrival time was 100ms -- deemed as a suitable granularity for periodic sensor readings. No matter the number of messages sent, the TtFDF remained constant for each architecture -- Figure \ref{scaleTfd} shows the results for 1 message: 

\begin{figure}[!htb]
\centering
\vspace{-10mm}
\includegraphics[width=3.45in]{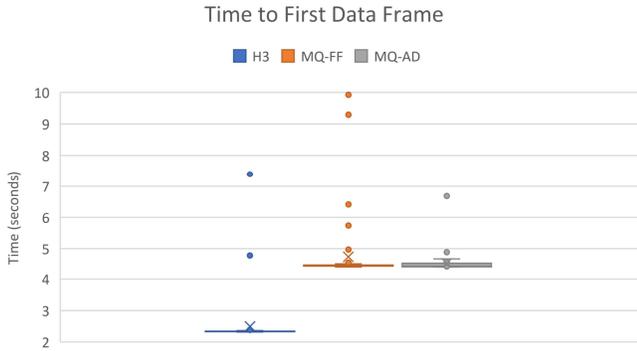}
\vspace{-16mm}
\caption{Time to First Data Frame Comparison (1 Message)}
\label{scaleTfd}
\end{figure}

\begin{figure}[!htb]
\centering
\vspace{-10mm}
\includegraphics[width=3.45in]{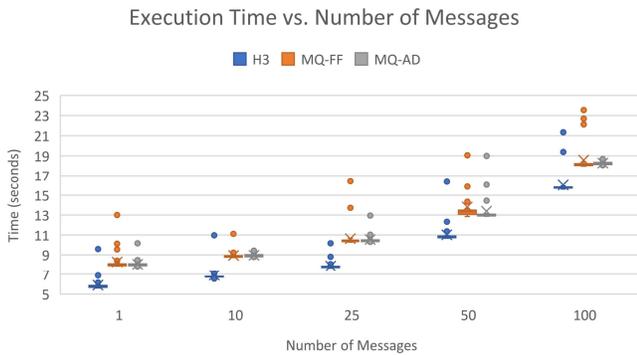}
\vspace{-16mm}
\caption{Load Scaling - Execution Time}
\label{scaleExc}
\end{figure}

MQ took an additional 2 seconds (1-RTT) to reach the first data frame. This was also seen in our previous work \cite{cscnpaper}. MQ uses this time for authentication with the broker, whereas H3's is included in the request header. The savings of 1-RTT is reflected in the execution time of the respective architectures as well, shown in Figure \ref{scaleExc}. Such a savings becomes more apparent the higher the RTT and, in this case, meant a difference of seconds in the dissemination of information.

\begin{figure}[!htb]
\centering
\vspace{-10mm}
\includegraphics[width=3.45in]{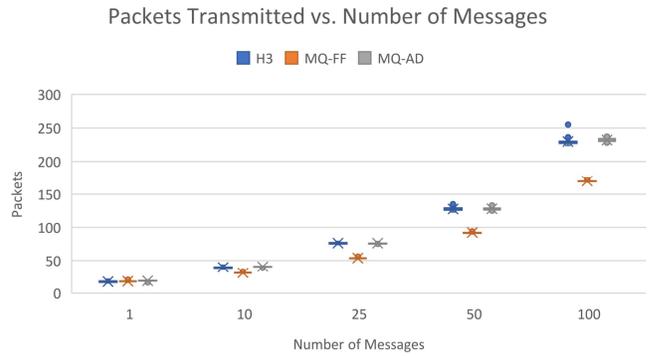}
\vspace{-16mm}
\caption{Load Scaling - Packets Transmitted}
\label{scalePkt}
\vspace{-3mm}
\end{figure}

\begin{figure}[!htb]
\centering
\vspace{-10mm}
\includegraphics[width=3.45in]{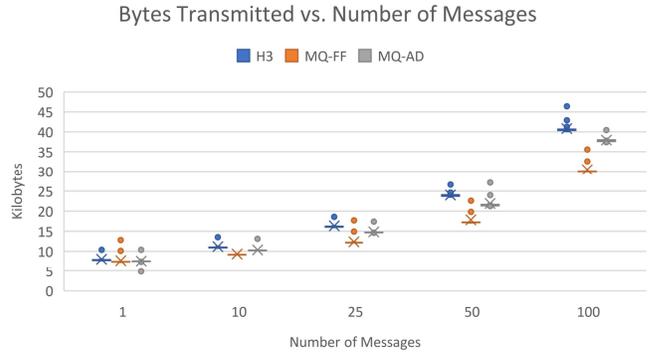}
\vspace{-16mm}
\caption{Load Scaling - Bytes Transmitted}
\vspace{-6mm}
\label{scaleByt}
\end{figure}

In terms of network overhead, packets transmitted are shown in Figure \ref{scalePkt}. It can be seen that, across the range of messages, H3 and MQ-AD generated roughly a third more packets than MQ-FF. There are two reasons for this: 1.) both H3 and MQ-AD messages were ACK-eliciting and 2.) neither MQ-FF nor MQ-AD make use of stream multiplexing and therefore do not generate MAX\_STREAM frames. The transmitted bytes, Figure \ref{scaleByt}, followed a similar trend as with packets transmitted for the same reasons.

\begin{figure}[!htb]
\centering
\vspace{-11mm}
\includegraphics[width=3.45in]{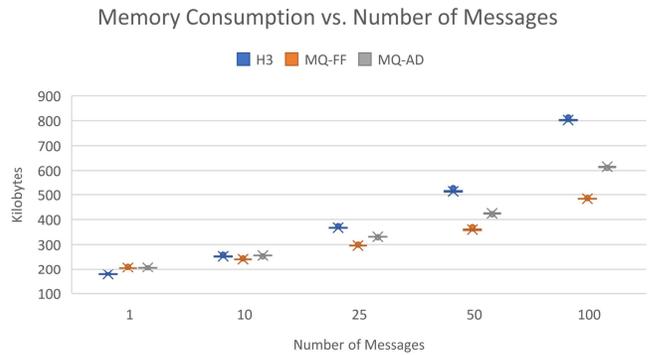}
\vspace{-16mm}
\caption{Load Scaling - Memory Consumption}
\label{scaleMem}
\vspace{-1mm}
\end{figure}

Memory consumption is shown in Figure \ref{scaleMem}. In the case of a single message, H3 consumed 28KB less memory than MQ-FF. As the number of messages increased however, a crossover point was reached. After looking through the profiler traces, this was found to be due to i.) H3's \textit{RequestWriter} buffer (discussed in Section V), and ii.) the fact that MQ's code has been written to only ever use one stream. The H3 implementation created a new stream ID per request, which accounted for 75KB of additional memory for 100 messages.

\begin{figure}[!htb]
\centering
\vspace{-10mm}
\includegraphics[width=3.45in]{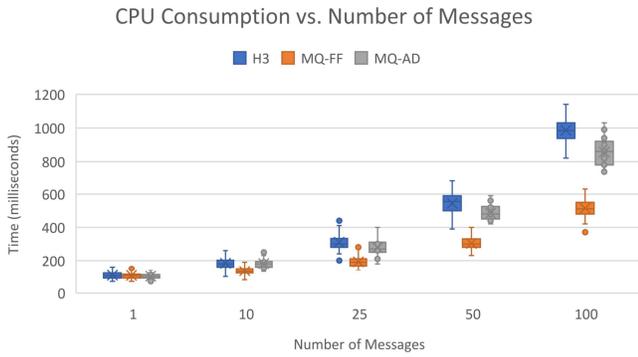}
\vspace{-16mm}
\caption{Load Scaling - CPU Consumption}
\label{scaleCpu}
\vspace{-3mm}
\end{figure}

In Figure \ref{scaleCpu}, H3 was found once again to have a higher overhead on the device. H3's QPACK decoding was identified as a sizable CPU consumer. H3 also exhibited noticeably more variability than MQ variants, indicated by the height of its whiskers. While the median and mean values of H3's CPU consumption were higher, overlap still remained when comparing the zones occupied by each method's whiskers.

\begin{figure}[!htb]
\centering
\vspace{-10mm}
\includegraphics[width=3.45in]{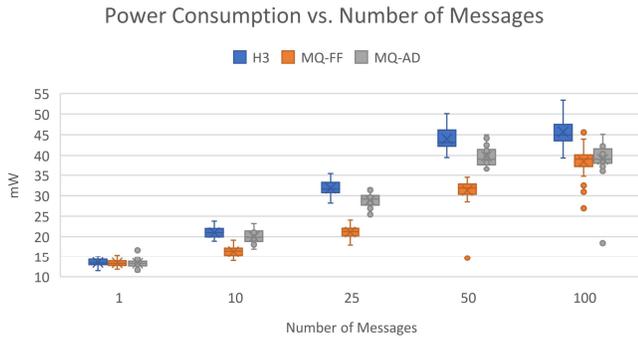}
\vspace{-16mm}
\caption{Load Scaling - Power Consumption}
\label{scalePwr}
\end{figure}

Shown in Figure \ref{scalePwr}, MQ consumed less power, although H3 remained competitive. Given H3 and MQ-AD's additional data transmission overhead, as well as H3's CPU and memory consumption, it was not surprising that MQ-FF used the least power. While H3 experienced more variability, MQ-FF and MQ-AD seemed to produce more outliers at higher scale.

\subsection{Effects of RTT}
We also investigated the RTT's impact on the metrics of interest. For 50 32-byte messages with an inter-arrival time of 100ms, the NetEm delay profile between publisher and broker was increased from 0 to 2 seconds in increments of 500ms. All delays followed a uniform distribution and NB-IoT rate throttling was applied throughout the testing.

\begin{figure}[!htb]
\centering
\vspace{-10mm}
\includegraphics[width=3.45in]{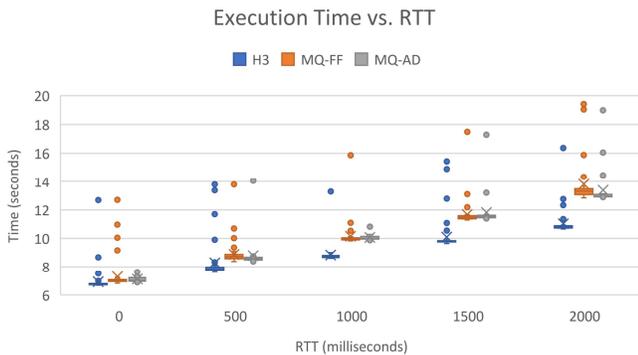}
\vspace{-16mm}
\caption{RTT Effects - Execution Time}
\label{rttExc}
\end{figure}

Figure \ref{rttExc} displays the effects of RTT on the execution time. With no delay imposed by NetEm, it can be seen that H3, MQ-AD, and MQ-FF all finished roughly after 7 seconds, with several outliers (barring MQ-AD). As the RTT increased however, the execution time, indicating performance, favored H3. This is explained by the previously mentioned overhead of MQ's authentication signaling, which is compliant with MQTT's protocol specification \cite{cscnpaper}.

\begin{figure}[!htb]
\centering
\vspace{-9mm}
\includegraphics[width=3.45in]{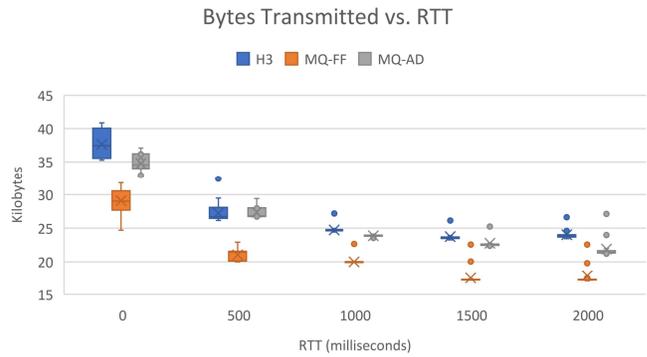}
\vspace{-15mm}
\caption{RTT Effects - Bytes Transmitted}
\label{rttByt}
\end{figure}

Figure \ref{rttByt} shows an interesting trend for each architecture's bytes transmitted. As the RTT increased, both the variability and the number of bytes transferred reduced. All methods reached a saturation point at 1500ms. This was due to the presence of PING frames padded with hundreds of bytes at lower RTTs. RFC 9000 notes such behavior as Path MTU probing. Again, H3 and MQ-AD exhibited 33\% more overhead.

\begin{figure}[!htb]
\centering
\vspace{-9mm}
\includegraphics[width=3.45in]{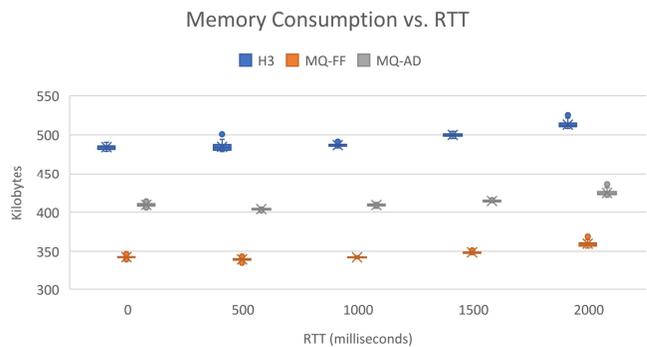}
\vspace{-15mm}
\caption{RTT Effects - Memory Consumption}
\label{rttMem}
\end{figure}

A modest increase in memory consumption was noted for H3 and both methods of MQ at higher RTTs in Figure \ref{rttMem}. From no NetEm delay to a delay of 2000ms, H3 used 28KB more memory and MQ variants used 18KB more, on average. 

Profiling showed that the additional memory was consumed by a \textit{sync.Pool} packet buffer in QUIC-GO's packet handler. Neither CPU nor power consumption were impacted by RTT. Therefore, they are not plotted. We note that, compared to H3, MQ-FF had an edge of 35\% less CPU and 29\% less power consumption, on average.

\newpage

\subsection{Effects of Message Size}

With NB-IoT network emulation, 50 messages (32, 2048, and 4096 bytes) with an inter-arrival time of 100ms were investigated. These values were chosen to be large enough to span multiple packets (1, 2, and 4, respectively). The purpose of this test was to explore how a pub-sub application's message payload may affect the performance and overhead.

\begin{figure}[!htb]
\centering
\vspace{-10mm}
\includegraphics[width=3.45in]{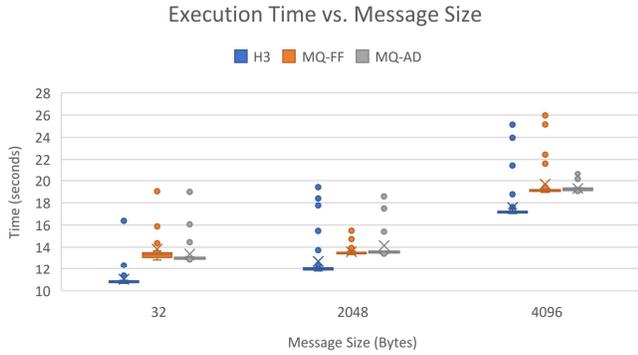}
\vspace{-16mm}
\caption{Message Size - Execution Time}
\label{messSizeExc}
\end{figure}

\begin{figure}[!htb]
\centering
\vspace{-10mm}
\includegraphics[width=3.45in]{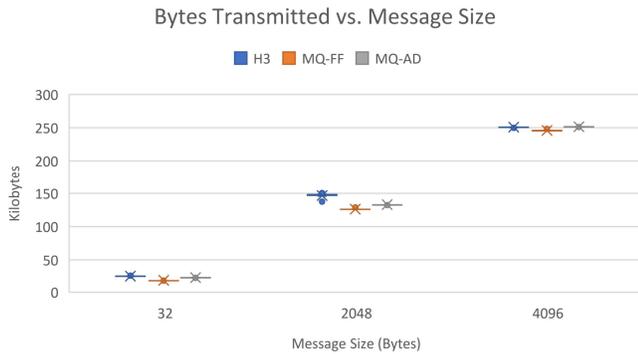}
\vspace{-16mm}
\caption{Message Size - Bytes Transmitted}
\label{messSizeByt}
\end{figure}

As shown in Figure \ref{messSizeExc}, the impact that message size had on execution time was comparable for each method. Moreover, in Figure \ref{messSizeByt}, the disparity of H3 and MQ-AD's network overhead became dwarfed by the increasing message sizes.

\begin{figure}[!htb]
\centering
\vspace{-10mm}
\includegraphics[width=3.45in]{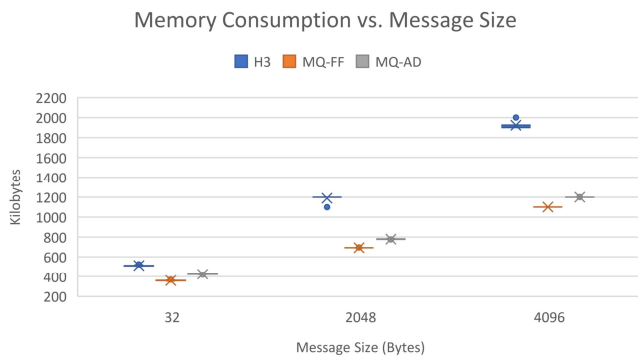}
\vspace{-16mm}
\caption{Message Size - Memory Consumption}
\label{messSizeMem}
\end{figure}

\newpage

\begin{figure}[!htb]
\centering
\vspace{-10mm}
\includegraphics[width=3.45in]{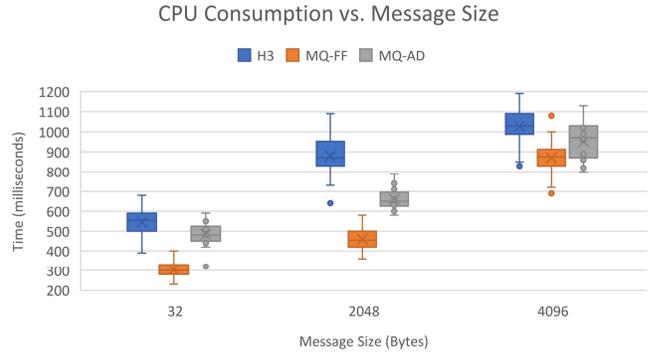}
\vspace{-16mm}
\caption{Message Size - CPU Consumption}
\label{messSizeCpu}
\end{figure}

\begin{figure}[!htb]
\centering
\vspace{-10mm}
\includegraphics[width=3.45in]{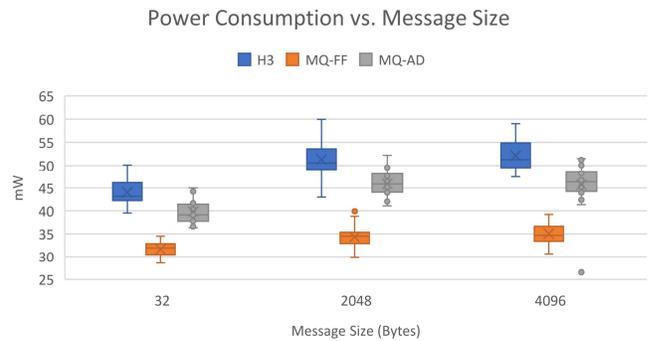}
\vspace{-16mm}
\caption{Message Size - Power Consumption}
\label{messSizePwr}
\end{figure}

H3's memory gap exacerbated in Figure \ref{messSizeMem}. The H3 \textit{RequestWriter} function attributed to 9.69\%, 30.75\%, and 49.84\% of the total memory consumed in the respective cases. Additional memory consumption by the packet buffer was common to both architectures. On the other hand, H3's CPU usage remained competitive in Figure \ref{messSizeCpu}: scheduling and the packet handler accounted for the majority of CPU increase for all methods. H3 exhibited slightly more CPU variability. Power consumption in Figure \ref{messSizePwr} showed modest growth.

\subsection{Effects of Packet Loss}
Because MQ was not coded to make use of QUIC's stream multiplexing feature, we hypothesized that HoLB would be an issue and that a clear edge would be given to H3. 50 messages with an inter-arrival time of 100ms were sent with a fixed message size of 2048 bytes. A larger message size was chosen in this test so that the message would span multiple packets.

NB-IoT network emulation parameters were used and NetEm rules were applied to drop outgoing packets on the client interface according to a uniform distribution. The amount of packet loss applied ranged from 0 to 6\% in increments of 2\%. We note that due to the nature of pub-sub, exchanges did not produce high volumes of packets. Because of this, we chose rather high packet loss percentages to ensure each iteration would experience some level of packet loss.

\begin{figure}[!htb]
\centering
\vspace{-10mm}
\includegraphics[width=3.45in]{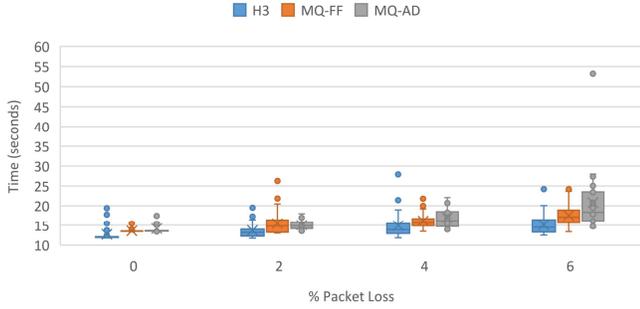}
\vspace{-14mm}
\caption{Packet Loss - Execution Time}
\label{lossExc}
\end{figure}

\newpage

As the amount of packet loss increased, execution times experienced much more variation -- extending the whiskers of each IQR. Extreme outliers were also observed in Figure \ref{lossExc} whereby packet loss had affected the handshaking process. In such instances, execution time took upwards of 25 seconds. It peaked at 60 seconds for MQ-AD with 6\% loss.

\begin{figure}[!htb]
\centering
\vspace{-10mm}
\includegraphics[width=3.45in]{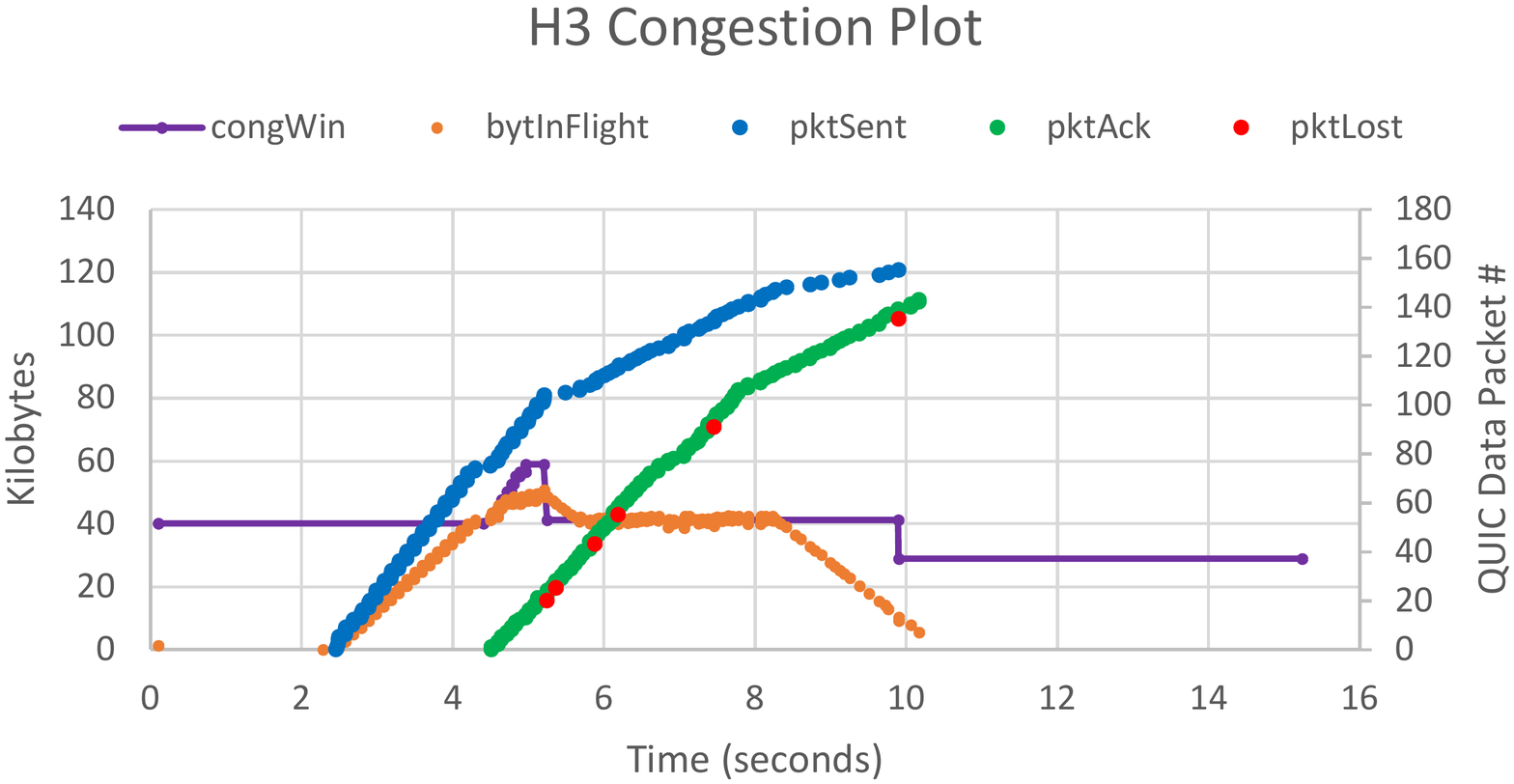}
\vspace{-16mm}
\caption{Packet Loss (4\%) - H3 Congestion Plot}
\label{h3Cong}
\end{figure}

\begin{figure}[!htb]
\centering
\vspace{-11mm}
\includegraphics[width=3.45in]{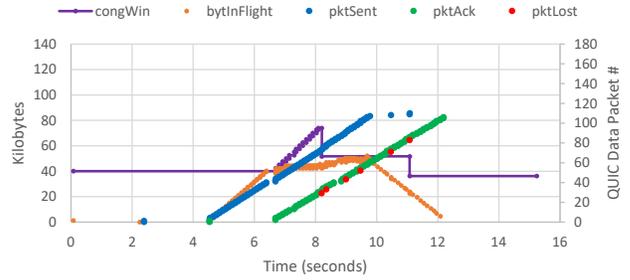}
\vspace{-18mm}
\caption{Packet Loss (4\%) - MQ-FF Congestion Plot}
\label{mqFfCong}
\vspace{-2mm}
\end{figure}

\begin{figure}[!htb]
\centering
\vspace{-10mm}
\includegraphics[width=3.45in]{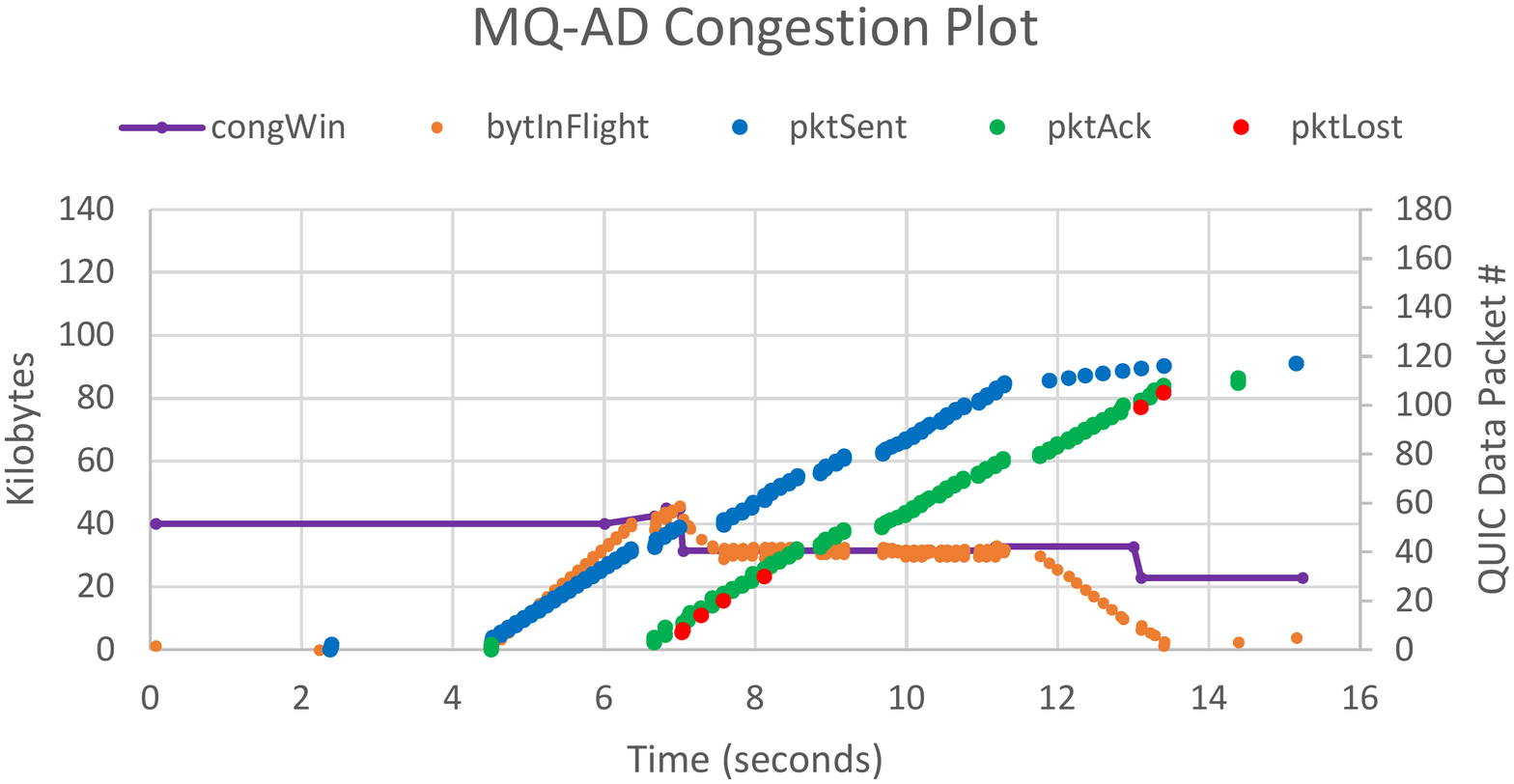}
\vspace{-18mm}
\caption{Packet Loss (4\%) - MQ-AD Congestion Plot}
\label{mqAdCong}
\vspace{-4mm}
\end{figure}

In order to further dissect HoLB, congestion plots were generated from qlog traces. These are shown for H3, MQ-FF, and MQ-AD in Figures \ref{h3Cong}, \ref{mqFfCong}, and \ref{mqAdCong} respectively. Traces were selected from the 4\% loss experiments -- there were 6 packet losses for H3, 6 for MQ-FF, and 7 for MQ-AD.

All methods had an initial congestion window of 40KB, following the left-hand vertical axis. Data transmission for MQ-AD and MQ-FF lagged H3 by 1-RTT. Small gaps in the \textit{Packets Sent} plots of H3 and MQ-AD were solely caused by the \textit{Congestion Window} being full. For these methods, loss events did not cause HoLB (even though MQ-FF only used a single QUIC stream). This was indicated by the absence of gaps in the \textit{Bytes in Flight} and \textit{Packets Acknowledged} plots.

On the other hand, MQ-AD's congestion plot \textit{was} indicative of HoLB in several spots, further explaining the more deteriorated MQ-AD execution time compared to other methods. At the 8.5, 9.2, 11.3, and 13.4 second marks, no data was sent or acknowledged and the bytes in flight were not updated giving way for about 2.3 seconds of blocking time. Loss, however, did not pose visible effects on memory, CPU, nor power consumption.

%%%%%%%%%%%%%%%%%%%%%%%%%%%%%%%%%%%%%%%%%%%%%%%%%%%%%%
\section{Conclusions}

In this paper, we aimed to provide a practical investigation into the considerations of tuning QUIC for IoT. We noted that RFC 9006 \cite{rfc9006} was written to ensure effective and resource-friendly behavior for TCP in IoT, as IoT greatly differs from traditional networks. To our knowledge, this paper is the first to provide equivalent insights for QUIC. We hope that these points prove be taken into consideration by researchers looking to adopt QUIC in their work with IoT.

We compared our HTTP/3 pub-sub application to the literature's state-of-the-art MQTT-over-QUIC in 2 different QoS modes. Different message payloads and network conditions were modeled in our study to show trends associated with the performance and resource consumption footprint for each of the studied architectures. While previous studies have looked at QUIC in IoT, HTTP/3 had not been previously considered. We hypothesized that, with its native integration of QUIC, HTTP/3 would perform better than the existing literature.

Indeed, we found that our H3 solution provided a performance savings of 1-RTT compared to MQTT-over-QUIC -- amounting to faster data delivery, on the scale of seconds. Our H3 application was also less susceptible to HoLB because of its use of stream multiplexing. On the other hand, the toll on the end device was slightly (yet consistently) higher than MQ-AD. This was true of memory, CPU, and power consumption.

Given these findings, we deem our H3 architecture to be preferable for higher powered devices and use cases where dissemination of information is time-sensitive and requires reliability. This is because of the lower latency and slightly higher resource consumption that our application provided.

QUIC parameters that we identified and tuned through our iterative process of code profiling, traffic captures, and qlogs have been outlined. For each of these points, our tuning methodology has been justified and the implications on performance and resource consumption have been detailed. Two novel contributions we made were our i.) high watermark scheme for MAX\_STREAM frames, reducing signalling, and ii.) connection-level basicAuth, reducing device CPU and memory requirements. Attention was also drawn to key QUIC timers and the \textit{max\_ack\_delay} to deal with larger RTTs.

One limitation of this work is its sole use of the QUIC-GO library. Many of the tuning details discussed are directly related to QUIC's RFC, however, and are expected to be common to all QUIC libraries and implementations. This work could be further extended by the investigation into additional QUIC libraries and implementations. Testing communications between a client running $QUIC Library_x$ against a server running $QUIC Library_y$ could also yield interesting results.

%%%%%%%%%%%%%%%%%%%%%%%%%%%%%%%%%%%%%%%%%%%%%%%%%%%%%%
\bibliographystyle{ieeetr}
\bibliography{references.bib}

\end{document}